# Design of tunable GHz-frequency optomechanical crystal resonators

**Hannes Pfeifer[1], Taofiq Paraïso[1], Leyun Zang[1] and Oskar Painter[2*]**

[1]*Max Planck Institute for the Science of Light, Günther-Scharowksy-Straße 1, 91058 Erlangen, Germany*
[2]*Institute for Quantum Information and Matter and Thomas J. Watson, Sr., Laboratory of Applied Physics, California Institute of Technology, Pasadena, USA*
[*]*opainter@caltech.edu*

**Abstract:** We present a silicon optomechanical nanobeam design with a dynamically tunable acoustic mode at 10.2 GHz. The resonance frequency can be shifted by 90 kHz/V$^2$ with an on-chip capacitor that was optimized to exert forces up to 1 μN at 10 V operation voltage. Optical resonance frequencies around 190 THz with Q factors up to $2.2 \times 10^6$ place the structure in the well-resolved sideband regime with vacuum optomechanical coupling rates up to $g_0/2\pi$ = 353 kHz. Tuning can be used, for instance, to overcome variation in the device-to-device acoustic resonance frequency due to fabrication errors, paving the way for optomechanical circuits consisting of arrays of optomechanical cavities.

## 1. Introduction

Within the last decade the field of cavity-optomechanics has achieved great progress in exploring and exploiting the interaction of light with mechanical motion [1], realizing laser cooling of a mechanical resonator to its quantum mechanical ground state [2,3] and generating both squeezing of light [4,5] and mechanical motion below the standard quantum limit [6,7]. The applications of cavity-optomechanical systems for fundamental and applied sciences are myriad, ranging from gravitational wave detection to sensing of inertial forces [8-11]. Coupled optomechanical oscillators were also proposed as a toolbox to study quantum many body dynamics and as a promising system for information processing and routing [12-18]. In this regard, optomechanical crystals which combine photonic and phononic band engineering in an optically thin device layer [19,20] offer great versatility to implement different regimes of cavity-optomechanics in a chip-scale, integrated platform.

A key role for the realization of arrays and networks of optomechanical cavities is the control over the exact properties of the single sites and their coupling [12-18]. For example small deviations in the mode frequencies can shift coupled cavities out of resonance with each other and prevent an efficient excitation transfer between them. In optomechanical crystals such variations are unavoidably introduced through fabrication non-idealities. The shifts of acoustic resonances that thereby arise are usually in the order of 0.1-0.01% of the target frequency for well-developed fabrication processes. Nevertheless a first demonstration of phononically coupled optomechanical cavities in an optomechanical crystal was recently presented using shared super modes of coupled cavities [21]. To compensate detuning one may use a static approach like post fabrication processing methods such as nano-oxidation shown to tune photonic cavities [22], but dynamic control over mechanical resonances on chip could extend the abilities to shape phonon storage and propagation [23].

Silicon optomechanical crystals are produced by patterning and releasing the device layer of a silicon on insulator (SOI) chip [19]. The periodic structuring allows the design of photonic and acoustic bandstructures governing the dispersion of the respective field in the medium [20]. Introduction of defects through a continuous deformation of the periodic unit cell can be used to localize both vibrational and optical modes. A mechanical mode periodically changes the refractive index distribution through vibrating boundaries and stress. A spatial overlap with an optical mode leads to a coupling of both modes similar to the canonical optomechanical model of a moving end mirror in a Fabry-Pérot cavity. Both 1D and 2D systems can be realized, where in the 1D case the light (vibration) is confined on a silicon nanobeam in the lateral direction through index contrast (material boundaries) and along the beam through a photonic (phononic) bandgap.

It has been shown that in the SOI chip-scale platform, dynamic tuning of optical modes can be achieved by a static displacement of the optical cavity structure induced through an on-chip capacitor [24,25]. For example, by applying a voltage across two suspended silicon beams which form an optical waveguide or cavity, changes in the gap between the beams can be used to sensitively tune the optical waveguide propagation phase [26] or cavity mode frequency [24].

The necessity to also tune acoustic resonators naturally arises with a growing ensemble of coupled, interacting oscillators, whether they be instruments in an orchestra or optomechanical cavities on a chip. The various techniques to shift acoustic resonators' frequencies vary of course with their size, but they usually all aim at a modification of the geometry or a change of the restoring force for example through induced stress or a change of the elasticity of the vibrating materials. Doing this for chip-scale oscillators is challenging and has been demonstrated for oscillators up to MHz frequencies, for example, of flexural nanobeam modes [27-30]. However the implementations of on-chip circuits and networks usually require a regime where the mechanical frequency has to be larger than the optical dissipation rate, the so-called resolved sideband regime, which can be reached for GHz-modes in optomechanical crystals.

Here we propose to use capacitive forces to tune the frequency of an acoustic GHz mode of an optomechanical crystal nanobeam structure. The principle is schematically depicted in Fig. 1(a) and resembles the tuning of a string through tension. However, instead of modifying a flexural mode's frequency, usually appearing at some MHz, we aim at tuning the localized phononic crystal modes which occur at GHz frequencies. These modes emerge from vibrations of the unit cells of the 1D acoustic crystal, where a localization is realized by surrounding cells featuring an acoustic bandgap at the frequency of the localized mode. A force exerted onto the nanobeam would stretch the unit cells and with that change their periodicity and thereby the vibrational frequency.

## 2. On-chip forces for tuning acoustics

A critical parameter in the proposed tuning scheme is the force that can be exerted by the capacitor onto the beam. Since silicon is a rather stiff material comparably large forces have to be exerted to stretch a silicon nanobeam by a reasonable fraction. On the other hand the parameters of the capacitor, such as the size and the operational voltage, have to be compatible with the implementation on chip. To roughly estimate the available forces one can assume a simple plate capacitor (see Fig. 1(a)) with the well-known result

$$F = -\frac{\partial}{\partial(d)}\left(\frac{C(d)}{2}U^2\right) = -\frac{\partial}{\partial(d)}\left(\frac{\varepsilon \cdot h \cdot L}{d}\right) \cdot \frac{U^2}{2} = \frac{\varepsilon \cdot h \cdot L}{2 \cdot d^2}U^2 . \quad (1)$$

For an air-filled gap ($\varepsilon = \varepsilon_0$) and realistic capacitor parameters such as a metal layer thickness $h$ of 150 nm, a length $L$ of 50 μm and a capacitor gap $d$ of 100 nm, one arrives at 3.3 nN/V² yielding about 330 nN for an applied voltage $U$ of 10 V.

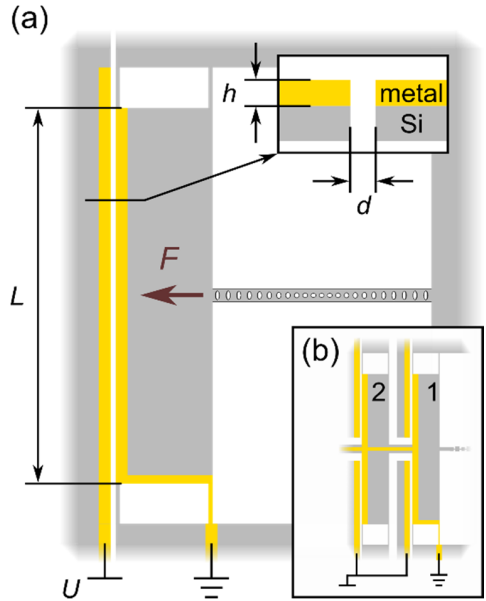

Fig. 1(a) Tuning principle using an on-chip capacitor that exerts a force onto an optomechanical crystal nanobeam. The force is determined by the capacitor geometry and the applied voltage. (b) The maximal force, given by the capacitor's breakdown voltage, can be increased by stacking capacitors on multiple connected pads.

To produce larger forces one may reduce the capacitor gap or increase the voltage, however it is known from measurements of the modified Paschen-curve for small capacitor gaps that for a certain setup the breakdown voltage of the capacitor decreases with smaller gaps [31-34]. A linear dependence would then decrease the maximal voltage and with that the maximal force by the same amount the gap size shrinks. An alternative approach to boost the exertable force that does not aim on an improvement of the single capacitor is a stacking of capacitors as shown in Fig. 1(b). There the silicon pads that support the capacitors are connected in the center. Also the movable capacitor plates can be connected via this line to share a common electric potential. Silicon teeth supporting the other sides of the capacitors can protrude into the space in between the silicon pads from the frame around. If a voltage is applied in this configuration, all capacitors pull together on the nanobeam like a team in a tug-of-war match. Since the capacitors pull on a rigid nanobeam that does not stretch much, the displacement still stays small. Stacks of this kind can be found in general textbooks on microelectromechanical system (MEMS) designs (such as [35]) and in MEMS applications that require strong forces for actuation or on-chip mechanical manipulation (e.g. [36,37]).

For a more rigorous evaluation of the extractable force the single capacitor and the stacked design were simulated with a finite element software package [38]. The force exerted into the direction of the attached nanobeam is found by integration of the respective traction vector component. The traction or stress vector $T$ giving the force density per area on an imagined surface within the material can be found by projecting the stress tensor $\sigma$ onto the normal vector $n$ of the surface so that its components are given by $T_l = n_k \cdot \sigma_{kl}$. The integration was carried out on a fixed surface at the right boundary of the simulation domain, where in the complete design the acoustic nanobeam would be attached. Fig. 2 shows the displacement along the beam direction and the stress in the structure for the case of four stacked plate capacitors. Note that the displacement of the capacitor hosting silicon pads is almost negligible. The result for the single capacitor yields a smaller force than estimated with the assumption of a plate capacitor. This stems from the obvious deviation of the geometry from an ideal plate capacitor with huge contributions of edge fields and from some fraction of the force (~1%) that is diverted to the tethers that hold the suspended silicon pads. The simulations show that a single capacitor with the same geometry as described before exerts a force of about 150 nN, if 10 V are applied. The silicon layer has a thickness of 220 nm and the supporting pad a length of 50 μm and width of 5 μm. The four silicon tethers that hold the suspended pad have a width of 150 nm and length of 5 μm. Adding more capacitors to a stacked design roughly adds up the forces that are exerted by the single capacitors. To reach a force of 1 μN with a voltage of 10 V a stack of 7 capacitors would hence be sufficient. The footprint of a capacitor configuration with 7 single pads would roughly be 60 μm × 60 μm.

This gives an idea of the magnitude of the forces that can be available with capacitors on a SOI-platform. To correct the frequencies of the localized acoustic modes by the required 0.1 to 0.01%, it is now crucial to optimize the sensitivity of the modes to an applied force.

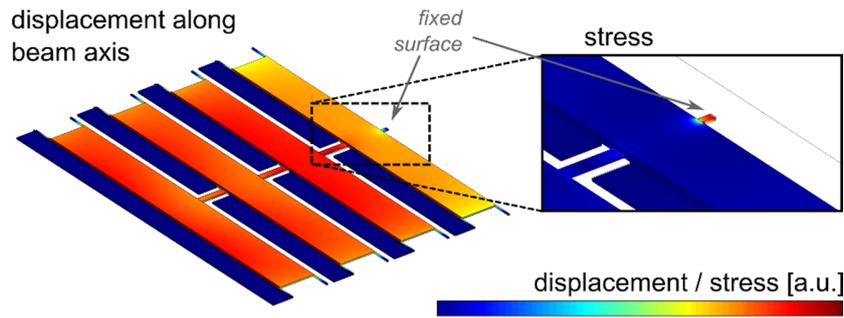

Fig. 2 Simulated [38] displacement along the beam axis and stress distribution in the structure for the case of 4 connected capacitors. All boundary surfaces and the unreleased capacitor plate are assumed to be fixed in this configuration. The maximum displacement of the capacitor carrying pads being hold by the connected beam is as low as ~5 pm/V$^2$. The largest stress and thereby the exerted force appears on the transition element towards the nanobeam.

## 3. Acoustic mode design

The most commonly used types of GHz vibrational modes on silicon nanobeams are so called breathing modes, where the unit cells expand and contract laterally in the beam [39,2]. Unfortunately the effect of an applied force to such a beam is almost negligible. This disadvantageous behavior has two reasons. First, this type of mode is in general not very sensitive to a stretching of the unit cell. It is created from an acoustic band with rather flat dispersion causing no significant frequency shift as the unit cell is stretched. Second, since the geometry was not optimized for this purpose the comparably large cross section of the wide beam makes it harder to stretch and therefore all of its mechanical modes are less sensitive to an applied force. Fig. 3(a) shows a unit cell of a common silicon nanobeam containing an elliptical hole. The smallest cross-section consists of two connection areas. If one assumes a fixed, smallest fabricable size for such a geometry feature, having two connecting parts means to double the smallest possible connection area and with that the stiffness of the beam. A more suitable design for this purpose is therefore a chain of connected rectangular silicon pads (see Fig. 3(a)). The minimum cross-section is smaller and due to its similarities with a basic spring-mass system it is easily customizable [20]. Adjusting the size of the pads and connecting bridges allows to modify the bandstructure and open large bandgaps. It is therefore an excellent candidate to search for modes with an initially higher force-sensitivity and to optimize their properties.

To test different types of modes for their sensitivity to an external force different unit cells were designed to exhibit large bandgaps in different frequency regimes. An example of such a structure is shown in Fig. 3(b). Since the unit cell is mirror symmetric about the xy- and the xz-planes, it is possible to classify the modes with respect to these symmetries. The resulting modes exhibit even or odd parity, upon the vector parity operations $\sigma_y$ (corresponding to the xz-plane) and $\sigma_z$ (xy-plane) that map the displacement field vector $(u(x,y,z),v(x,y,z),w(x,y,z))$ onto itself. For example an even $\sigma_y$ parity manifests as $\sigma_y(v(x,y,z)) = -v(x,-y,z)$ switching both the displacement and position vector component. The other two displacement components are not switched. The color highlighted modes in Fig. 3(b) correspond to even modes upon both symmetry operations. A full bandgap (highlighted in orange) opens above the first band of a symmetric mode (in blue). A deformation of such a mirror unit cell can now pull bands into the gap yielding a structure that supports a mode at the gaps frequencies. This deformed defect unit cell is now inserted into a chain of unperturbed cells with a large band gap. To smoothen the transition some more cells are added around. Their geometry sweeps continuously from the mirror to the defect cell. The resulting localized mode for the case of the first symmetric mode is shown in Fig. 3(c). Here the silicon pad of the defect unit cell is shrunk shifting its vibration to higher frequencies into the bandgap of the surrounding mirror cells.

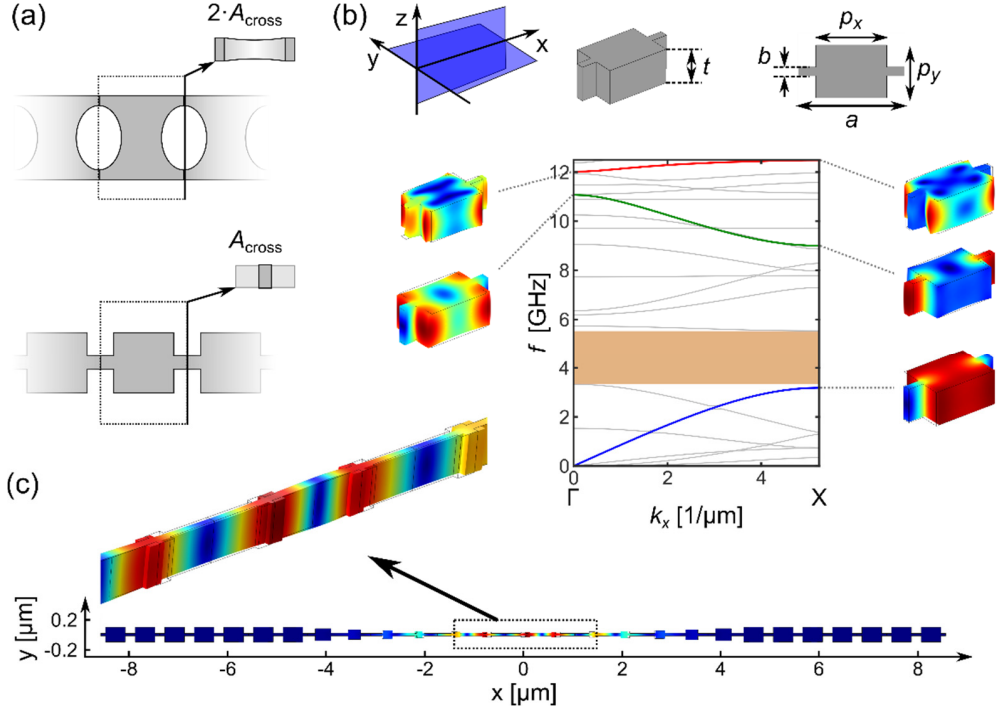

Fig. 3(a) Unit cells and cross-sectional area for different nanobeam designs. Smaller cross sections support potentially stronger tunabilities. (b) Geometry in the following $(b,p_x,p_y,a,t)$ = (50,400,300,600,220) nm and corresponding band diagram of the "connected-pad" design with a full bandgap. Highlighted are modes that are even under $\sigma_y$ and $\sigma_z$ vector parity operation (mirror symmetry in xy-/xz-plane). The modal plots show the magnitude of displacement at Γ- and X-point. (c) Displacement at a nanobeam featuring a localized acoustic mode in the bandgap of the outer cells from (b). The deformation of the central unit cells to $(b,p_x,p_y,a,t)$ = (50,100,100,640,220) nm pulls the mode's frequency of the first symmetric band of (b) into the gap. Simulations for (b) and (c) were performed with [38].

The described procedure was followed for modes constructed by the first four symmetric bands with appropriate mirror and defect unit cells. Within the simulations [38] the resulting nanobeams were furthermore exposed to a force stretching the geometry. Afterwards, the eigenmodes of this stretched - stressed geometries were again simulated to identify the effect of a fixed force onto the localized modes. The most sensitive constructed modes in this sweep were found to be formed from the first symmetric band (blue band) and in the full cavity structure shown in Fig. 3(c).

To enhance the tunability of the acoustic resonance different geometry parameter changes were investigated and design rules for an ideal tunable unit cell were identified. First the stretchable parts need to form a large fraction of the cell. This leads to rather long connectors of the pads compared to the unit cell size. Second the elongation can be optimized by making the connectors lateral cross section small and third the stress/strain field of the mode needs to extend to the cell boundaries. The last point heuristically ensures a good coupling between adjacent cells and with that a decent group velocity of the waveguide mode. A good force sensitivity for a localized vibration is produced if these rules are applied to the design of the defect region. For the design introduced in Fig. 3 this results in a comparably small pad in the center with long bridges connecting its neighbors. Fig. 4(b) shows the mirror bandstructure and a detail plot with the exaggerated displacement field for such a nanobeam. This geometry choice leads to a tuning of 0.08 % of its unstressed frequency if a force of one micro Newton (μN) is applied, meaning that the target tuning of the previously mentioned 0.1-0.01% can be achieved.

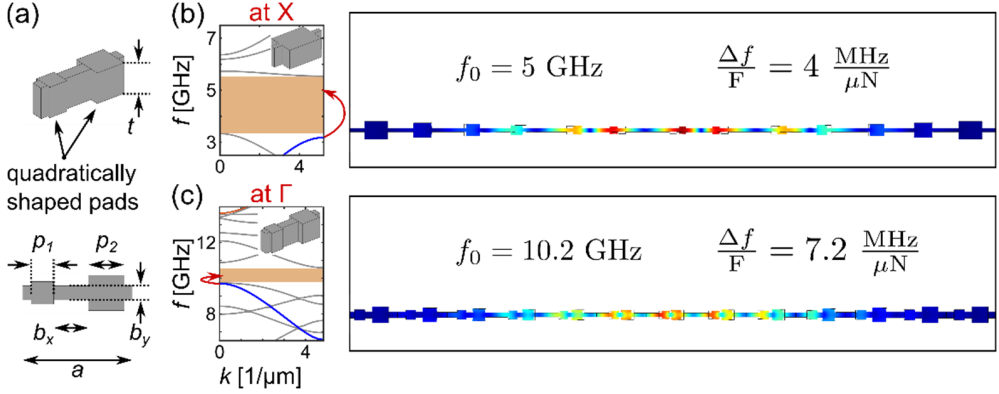

Fig. 4 Geometry and band diagrams of two different mirror unit cell designs, displacements of localized modes on acoustic nanobeams and their corresponding tunabilities. (a), definition of the geometry parameters for the new unit cell type used in (c). Design (b) corresponds to the geometry of Fig. 3, (c) employs a dimerization of the unit cell to pull a band of similar mode shape from the Γ-point. The mirror unit cell of (c) uses $(b_x, b_y, p_1, p_2, a, t)$ = (205,85,135,210,650,220) nm. The central cell uses $(b_x, b_y, p_1, p_2, a, t)$ = (325,50,100,125,650,220) nm. The nanobeam modes were designed through unit cell deformation and the tunabilities were extracted from simulations of the eigenfrequencies of differently pre-stressed structures.

Here the first symmetric band at its X-point was used to create localized modes. This causes the displacement field to exhibit a π-phase shift from one unit cell to the next. As detailed in [20] and later in the text this may be an unfortunate choice to ensure a large optomechanical coupling. It is therefore important to note that this mode design and its optimizations can also be used to create a localized mode from the same band at the Γ-point. In order to do this we utilized a dimerized unit cell as discussed in Ref. [40]. The mode shape of silicon pads oscillating against each other is thereby maintained, but instead of having equally sized pads, two different sizes are now employed. To ensure its usability also for photonic bands around 1550 nm, the lattice constant is kept similar (650 nm). Putting two pads into a unit cell of this size therefore shifts the mode of the two pads oscillation against each other to about twice the frequency of the previous design. Fig. 4(c) shows an example for such a design from a 220 nm silicon film where one can see the unit cells around the central defect oscillating in phase as expected for their Γ-point. The tunability remains large at 0.071% per 1 μN.

Up to now we have just considered localized acoustic modes. In the following, optical modes and their coupling to the mechanical field are optimized to yield tunable optomechanical nanobeams.

## 4. Optical mode design

The acoustic nanobeam design has thus far been optimized solely for mechanical tunability. The nanobeam structure must also support a localized optical resonance in order to create large radiation pressure coupling between light and the acoustic resonance of the beam. Because of its thin beam design, required for significant tuning of the acoustics, near-IR light in the 1500 nm telecom band will not be well guided in the acoustic nanobeam of the previous section. The lack of larger domains of silicon with its high refractive index, shifts all optical bands to higher frequencies [41]. Looking at the photonic bandstructure of the acoustic beam's defect cell, as shown if Fig. 5(a), this manifests in bands that are very close to the light cone. Usually one would try to use photonic bands at the X-point and a bandgap to localize an optical mode in a similar fashion as used in previous nanobeam optomechanical crystal cavities [39,20]. This cannot efficiently be done with this kind of unit cell, since the vicinity of the light cone would cause scattering to freespace modes and therefore lead to small optical quality factors.

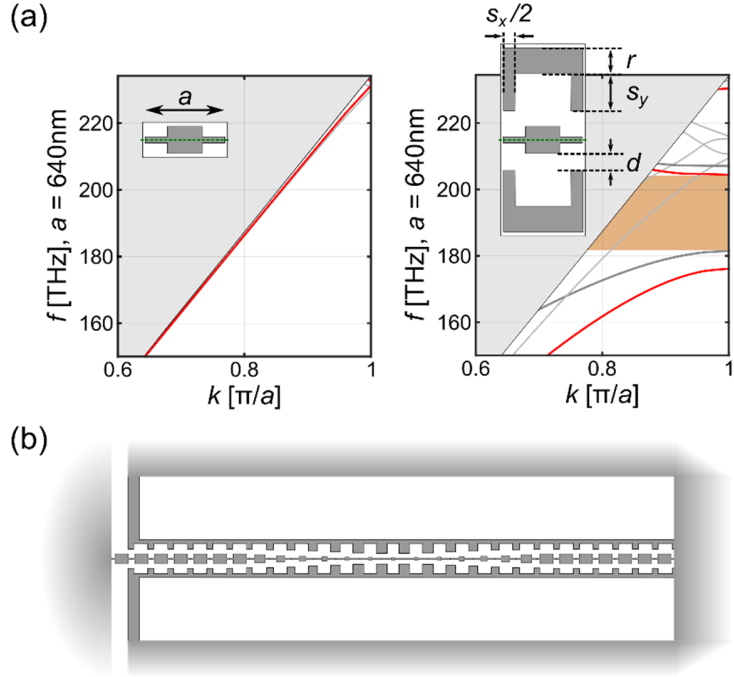

Fig. 5(a) Photonic bandstructure of a bare acoustic cell with $(b,p_x,p_y,a,t)$ = (50,280,220,640,220) nm and of the same cell with an added ribbon structure (simulated with [42]). The geometrical parameters used here are $(s_x,s_y,r,d)$ = (192,296,211,130) nm. The increasing amount of high refractive index material pushes the photonic bands towards lower frequencies. TM-like modes are plotted in light grey, TE-like modes appear in red (antisymmetric to green dashed plane) and dark grey (symmetric). (b) Proposed structure geometry, so that the added ribbons cannot affect acoustic properties or the force applied central beam.

In order not to affect the mechanical properties of the central acoustic beam, we choose here to add more high index silicon to the unit cell in the form of two additional "ribbons", which are mechanically disconnected from the central beam. Fig. 5(b) shows a schematic of such a beam structure where the central acoustic beam is accompanied on each side by a ribbon. The ribbons can be attached to the surrounding silicon on both ends while the central beam is connected to the capacitor carrying pad at one side to exert a force on it. Since the ribbons are made out of the same silicon device layer, they have the same thickness as the central beam. Also, they should be close enough to the center to form a single photonic waveguide with the acoustic part. To enhance the design capabilities for photonic bandstructures and modes, the inner edge of each ribbon can be altered by adding a silicon knob in the unit cell. As a result the photonic bands (simulated with "MIT photonic bands" software package [42]) are pushed away from the light cone to lower frequencies. By shaping the added knobs, large bandgaps for fields of certain symmetries can be obtained as shown in Fig. 5(a) (TE-like modes bandgap). These features enable one to independently optimize optical properties while keeping the mechanics unchanged.

Using the geometry sketched in Fig. 5(b) it is possible to design nanobeams with mechanical and optical modes, where the optical quality factor can exceed $10^6$ as extracted from FEM simulations [38]. As already mentioned those beams would not exhibit a large optomechanical coupling due to the initial choice of an X-point mechanical mode. In an infinite waveguide this choice manifests in a $\pi$ phase shift of the mechanical displacement field between adjacent unit cells. Any coupling in a single cell appearing as a frequency shift of the optical mode during the acoustic oscillation is therefore cancelled by a shift of opposite sign in the next cell. In general the linear coupling $g$ can be expressed in first order perturbation theory as a change of the optical modes frequency through a displacement-induced change of the refractive index distribution. For silicon photonics the major contributions originate from two different effects. First the change of the geometry through the displaced boundaries (MB) as detailed in [43,39] and second the stress induced refractive index change through the photo-elastic effect (PE) in silicon [44,39].

The moving-boundary contribution can be written as follow

$$g_{0,MB} = -\frac{\omega_0}{2} \cdot \frac{\int_{\delta V} (\vec{Q} \cdot \vec{n}) \cdot (\Delta\varepsilon \vec{E}_\parallel^2 - \Delta\varepsilon^{-1} \vec{D}_\perp^2) \, dS}{\int \vec{E} \cdot \vec{D} \, dV} \cdot x_{zpf} \quad (2)$$

Where $Q$ is the normalized displacement field (max$\{Q\} = 1$), $\Delta\varepsilon = \varepsilon_{silicon} - \varepsilon_{air}$ is the permittivity difference and $\Delta\varepsilon^{-1} = \varepsilon_{silicon}^{-1} - \varepsilon_{air}^{-1}$ the difference of the inverse permittivity. Note that the scalar product of the displacement field selects components that are normal to the material surface. $x_{zpf}$ is the zero point motion of the mechanical mode and hence $g_0$ the vacuum optomechanical coupling rate. The photo-elastic coupling can be calculated via

$$g_{0,PE} = -\frac{\omega_0}{2} \cdot \frac{\langle \vec{E} | \frac{\partial \varepsilon}{\partial \alpha} | \vec{E} \rangle}{\int \vec{E} \cdot \vec{D} \, dV} \cdot x_{zpf} \quad (3)$$

where the permittivity is changed through the generalized amplitude $\alpha$ of the vibration mode (with max$\{\alpha\} = 1$). In case of the photo-elastic effect it is given by

$$\frac{\partial \varepsilon_{ij}}{\partial \alpha} = -\varepsilon_0 n^4 p_{ijkl} S_{kl} \quad (4)$$

where $n$ is the refractive index of the material, $p_{ijkl}$ are the coefficients of the photoelastic tensor and $S_{kl}$ are the coefficients of the strain tensor of the acoustic mode.

In the structure shown here the largest optical field magnitude arises in the gap between the outer ribbons and the central acoustic beam. Therefore the photoelastic effect does not have a large contribution as the stress induced by the movement of the pads mainly appears in the connector parts where no large optical field is present. On the other hand the boundary movement of the acoustic beam that is created by the vibration changes the gap and therefore has a larger contribution. In case of the dimerized unit cell shown in Fig. 6 this means that the silicon pads in the center vibrate towards or away from the field maxima around the ribbons knobs depending on the phase of the oscillation. The employed acoustic mode here is a Γ-point mode, which means that in an infinite chain all unit cells vibrate in phase and therefore add up their contributions constructively to the total optomechanical coupling. The maximal values that can be reached in this configuration appear with minimal distances between the ribbon and the central parts. The example in Fig. 6(b) has a gap width $d$ (see Fig. 5(a)) of 50 nm, which realizes a waveguide vacuum optomechanical coupling rate $g_\Delta/2\pi = 467$ kHz from the moving boundary coupling for the X-point of optical modes of the first red highlighted band.

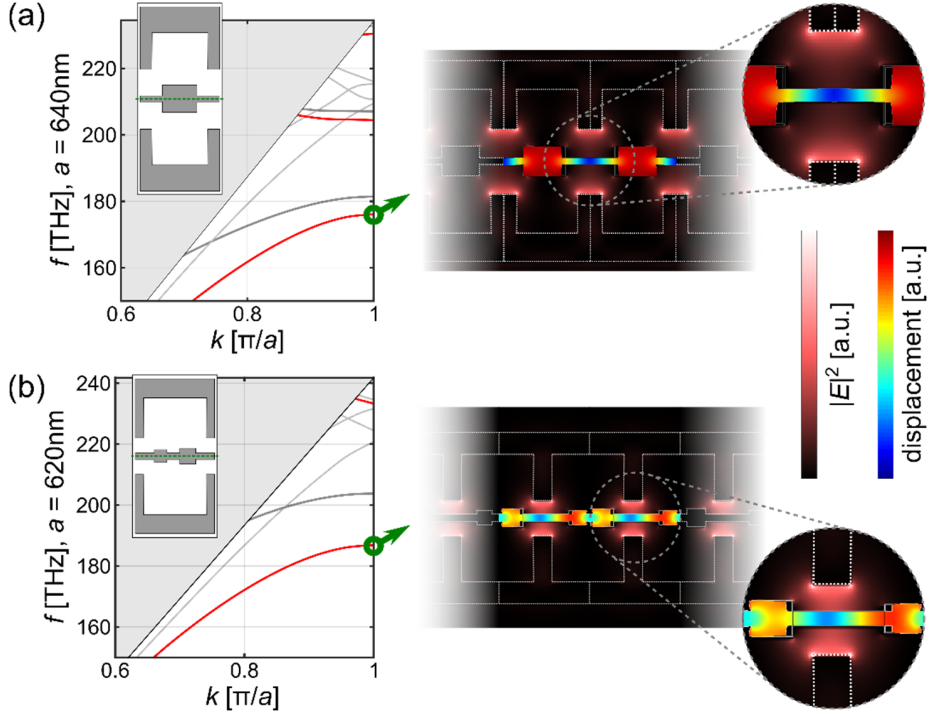

Fig. 6 Comparison of the photonic bandstructures and modes [42] of the single pad (a) and dimerized (b) defect unit cells with ribbons. The right column shows an overlay of the $|E|^2$ field magnitude at the X-point and the displacement field of the employed acoustic modes at the X- (a) and Γ-point (b) for a waveguide of the respective cells. In (a) the silicon boundaries along the waveguide are alternatingly displaced towards and away from the light field maxima. In contrast to that, the boundaries in (b) are along the whole waveguide either displaced away or towards the next light field maximum for a certain time in the oscillation cycle. The geometry of (a) is the same as in Fig. 5(a), while (b) uses $(b_x, b_y, p_1, p_2, a, t)$ = (295,50,100,125,620,220) nm and $(s_x, s_y, r, d)$ = (126, 317,150,50) nm.

To form an optical cavity the unit cell geometry of the defect introduced in Fig. 6(b) is swept from the center to that of mirror cells at the ends of the beam. This sweep is depicted in Fig. 7(a). The optical modes that can be found on the patterned nanobeam can be separated into different symmetry groups: (i) even and odd modes about the $\sigma_z$ mirror plane and (ii) even and odd modes about the $\sigma_y$ mirror plane. The definition of the planes and axis is the same as in Fig. 3(b). The dashed green line indicates the xz-plane (mirror plane for $\sigma_y$ operation). Note that we use a vector symmetry convention here. The fundamental TE-like optical modes of the silicon slab have $\sigma_z$ = +1. The fundamental transverse guided modes of the nanobeam have odd vector symmetry $\sigma_y$ = -1 (but even spatial symmetry). We will thus be interested in optical waveguide modes of $(\sigma_y, \sigma_z)$ = (-1, +1) vector symmetry.

The mirror unit cells band diagram plotted on the left of Fig. 7(a) features a large bandgap from about 165 THz to about 195 THz for TE-like modes of $\sigma_y$ = -1 symmetry. On the right are the band diagram and unit cell from Fig. 6(b). The plot between both shows the trace of the X-point frequencies along a quadratic deformation of one cell towards the other, where the defect cell is placed in the center of the sweep. The quadratic modulation of the geometry parameters thereby causes the band edge frequencies to roughly follow a harmonic potential. The fully assembled structure and a simulation of its optical properties yields an optomechanical cavity with an optical mode shown in Fig. 7(b). The Q factor of this resonance is simulated to be about $1.4\times10^6$ at a resonance frequency of 186 THz. The vacuum coupling rate to the localized acoustic dimer mode is determined to be around $g_0/2\pi$ = 291 kHz. Table 1 summarizes the parameters of this structure.

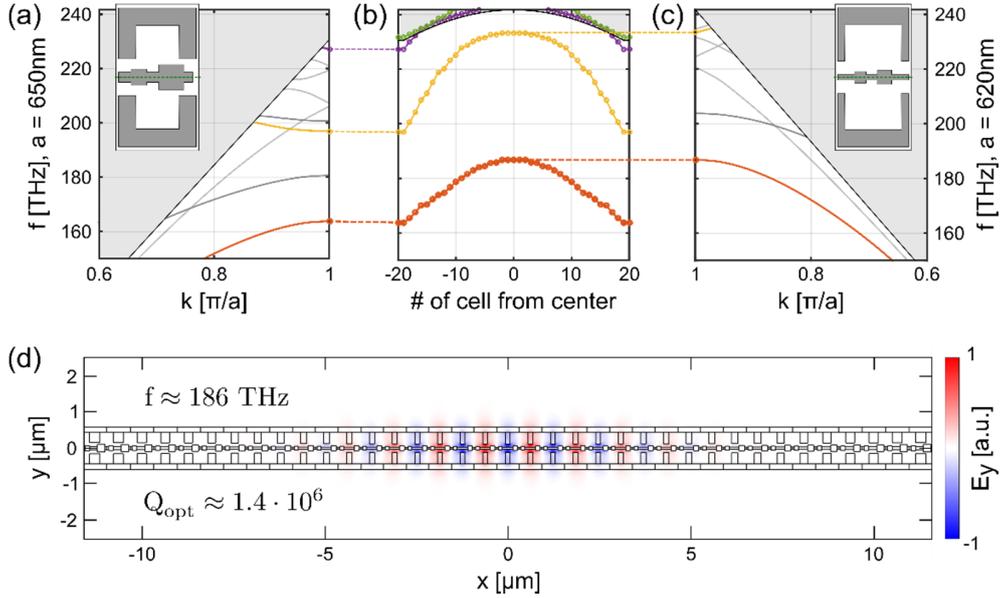

Fig. 7(a) Photonic bandstructure of dimerized mirror unit cell, (b) frequencies of the photonic X-point modes for the unit cell geometries along the nanobeam and (c) photonic bandstructure of the defect cell. (d) Plot of the $E_y$ component of the localized optical mode created from the first band by the deformation of unit cells.

The properties of the designed nanobeam can now be further optimized. To operate closer to the conventional communication band the unit cell size is reduced by shrinking the largest connector part. Thereby the tunability of the acoustic mode is maintained and the optical modes are shifted to higher frequencies. Table 1 also summarizes the parameters of some optimized structures closer to 190 THz and a design with a gap of 75 nm between the central acoustic beam and the ribbons that is less challenging to fabricate. The optimized structures were found by applying a Nelder-Mead simplex optimization onto some geometry parameters where the minimized cost function was chosen proportional to $-g_0^2 Q$ and contained a weighting dependent on the deviation from the target operation wavelength. The Q factor optimum was limited to 2 Million to bias it against unrealistic values.

**Table 1 Simulated nanobeam parameters**

| Nanobeam parameters | | 50 nm gap | | 75 nm gap |
| --- | --- | --- | --- | --- |
| | | Fig. 7 | optimized | optimized |
| Optics: | $f_{opt}$ | 186 THz | 190 THz | 190 THz |
| | $Q_{opt}$ | $1.4 \cdot 10^6$ | $2 \cdot 10^6$ | $2.2 \cdot 10^6$ |
| Mechanics: | $f_{mech}$ | 10.6 GHz | 10.8 GHz | 10.8 GHz |
| | $m_{eff}$ | 83 fg | 80 fg | 80 fg |
| | $x_{zpf}$ | 12 fm | 12 fm | 12 fm |
| Coupling: | $g_0/2\pi$ | 291 kHz | 353 kHz | 248 kHz |
| Mech. Tuning: | $\Delta f_{mech}/F$ | 9 MHz/μN | 9 MHz/μN | 9 MHz/μN |

## 5. Conclusion

We have presented a new design for on-chip SOI based optomechanical nanobeams with optical cavity modes in the 1500nm telecom band and tunable acoustic resonances in the microwave X-band (~ 10 GHz). Numerical simulations show that the device can operate in the well resolved sideband limit with optomechanical coupling strengths as large as $g_0/2\pi$ = 353 kHz. The tunability of the X-band mechanical resonances is up to 0.1% of the center frequency for micro-Newton tensioning forces, which can be realized with an applied voltage of 10V across a series of pulling capacitors. The ability to statically trim the acoustic properties of such microwave frequency acoustic devices in the setting of cavity-optomechanics would enable experimentation with arrays of coupled optomechanical devices, where for example one can explore transitions in the synchronization of optomechanical oscillators [45], slow and stop light in coupled resonator chains [46,23] create synthetic gauge fields for photons [16], or even study topological phases of sound and light [18]. We also anticipate that tunable or trimable optomechanical crystal structures will be valuable in creating optomechanical circuits capable of advanced signal processing in the framework of microwave photonics, such as narrowband filtering, amplification, and non-reciprocal signal transport [21].

## Acknowledgements

The authors would like to thank Matthew H. Matheny for valuable discussions on the benefits of dimerization of acoustic modes in phononic crystals. This work was supported by the AFOSR Hybrid Nanophotonics MURI, the Alexander von Humboldt Foundation, the Max Planck Society, and the Kavli Nanoscience Institute at Caltech.